\pdfoutput=1
\documentclass[11pt,a4paper]{article}
\usepackage[T1]{fontenc}
\usepackage[utf8]{inputenc}
\usepackage{amsmath,amssymb,amsthm,mathtools}
\usepackage{booktabs,xcolor,graphicx,float,needspace}
\usepackage{cite}
\usepackage[margin=2.6cm]{geometry}
\usepackage[colorlinks=true,linkcolor=blue!50!black,citecolor=blue!50!black,urlcolor=blue!50!black]{hyperref}

\numberwithin{equation}{section}
\newtheorem{proposition}{Proposition}[section]

\theoremstyle{remark}

\newcommand{\bx}{\mathbf{x}}
\newcommand{\bT}{\mathbf{T}}
\newcommand{\bA}{\mathbf{A}}

\title{\textbf{Spin--Torsion Coupling:}\\[4pt]
\large\textbf{Vector and Scalar Torsional Potentials of Chiral Spinor Currents}}
\author{Elisa Varani\\[2pt]\small Universit\`a Cattolica del Sacro Cuore, Italy}
\date{}

\begin{document}
\maketitle

\begin{abstract}
We investigate integral representations of torsional fields sourced by
chiral spinor currents within a weak-field, Ricci-flow-inspired
geometric framework. Starting from the coupling between spinorial
currents and axial torsion, we construct screened Green-kernel
solutions and examine their relation to antisymmetric tetradic
perturbations and contorsion.

Particular attention is devoted to stationary toroidal configurations,
for which the magnetic-type dipole moment vanishes while a nonzero
toroidal moment and effective circulation remain. The dependence of
these quantities on Yukawa screening is analyzed, together with their
implications for an Aharonov--Bohm-like geometric structure.

We further introduce a mixed-chirality source describing an effective
connection between oppositely polarized domains and investigate the
corresponding flip-sector integral. The dynamics of massive Majorana
neutrinos in an axial background is then examined, distinguishing
chiral mixing from helicity oscillations and identifying the role of
transverse torsional components.

The resulting framework connects spinorial sources, finite-range
torsional potentials, toroidal geometry, and neutrino spin dynamics,
while providing a mathematical basis for further investigations of
torsional transport and observable effects.
\end{abstract}

\section{Introduction}

In Refs.~\cite{r1,r2,r3}, stationary torsional configurations sourced by chiral
neutrino currents were described by integrals of the form
\[
\text{spinor bilinear}\ \longrightarrow\ \text{Green kernel}\
\longrightarrow\ \text{tensor field}.
\]
These representations involve left- and right-handed currents and a
mixed-chirality (flip) bilinear. The present work develops the
integral representations used in the Ricci-flow model of Ref.~\cite{r2}
and examines the conditions under which they can be used to reconstruct
the antisymmetric part of the tetrad perturbation. Each integral is a
functional of the spinorial source in the sense that it assigns a tensor
field to a given spinor configuration. The Einstein--Cartan spinor
framework underlying these sources is reviewed in~\cite{varani2025}.

The broader mathematical setting of integral representations of field
equations includes Bulnes's work on generalized Penrose transforms and
the cohomological approach to integral geometry developed by Bulnes and
Goborov~\cite{bulnes2013,bulnesgoborov2014}. Related work includes
hypercohomology and the Hodge decomposition of a
connection~\cite{bulnes2026}, as well as the monograph on
torsion~\cite{bulnes2022}. Here we use Green kernels acting directly
on spinorial sources.

The central question is how localized chiral spinor sources determine
the spatial structure and circulation of effective torsional fields.
Toroidal configurations provide a particularly useful setting in which
the geometry of the source, the screened Green response, and the global
circulation of the resulting potential can be examined separately. The
mixed-chirality sector then introduces a further question: how these
stationary structures may be related to the spinorial dynamics of
massive neutrinos in an axial background.

Section~\ref{sec:flow} recalls the Ricci flow and the stationary
topological configurations of Refs.~\cite{r1,r2}. Starting from the
Dirac Lagrangian, it expresses the axial torsion and contorsion through
the antisymmetric tetrad perturbation and writes the torsional part of
the modified Ricci flow in terms of it. Section~\ref{sec:yukawa}
introduces the screened axial potential, and
Section~\ref{sec:published} presents the three integral
representations of the antisymmetric tetrad sector.

Section~\ref{sec:torus} shows that the two current integrals contain
the same vector potential and computes its circulation on the
toroidal configuration, which is the stationary configuration of the
Ricci flow obtained in Ref.~\cite{r1}, modelled by a poloidal current
on a finite torus, with and without Yukawa screening.
Section~\ref{sec:fluct} turns to the mixed-chirality source. It
examines the role of the fermion mass and of uniform axial
backgrounds, and then computes the scalar potential of the flip
integral and its response at a planar interface between domains of
opposite chiral dominance.

\paragraph{Conventions.} We use the signature $(+,-,-,-)$ of \cite{r1,r2,r3}, with
$\Box=\partial_t^2-\nabla^2$, so that in the static limit $\Box\to-\nabla^2$. Latin indices $a,b,c$ are tetrad indices; at linear order they coincide with coordinate indices, and from Section~\ref{sec:published} on $a,b,c,k$ are spatial. All statements are at linear order in the
perturbations.

\section{Ricci flow, torsion and the antisymmetric tetrad sector}\label{sec:flow}

\subsection{Ricci flow and stationary topological configurations}

We retain the tetradic perturbation employed in the preceding studies.
To first order,
\begin{equation}\label{eq:metric}
e^a{}_{\mu}=\delta^a_\mu+h^a{}_{\mu},\qquad
g_{\mu\nu}=\eta_{\mu\nu}+2h_{(\mu\nu)}.
\end{equation}
The antisymmetric part $h_{[\mu\nu]}$ parametrizes an infinitesimal
local Lorentz change of the tetrad and leaves the metric unchanged at
this order.

Following Refs.~\cite{r1,r2}, the tetrad perturbation relaxes
according to
\begin{equation}\label{eq:flow}
\partial_\tau h_{\mu\nu}=-2R_{\mu\nu}(\Gamma),
\end{equation}
a model ansatz inspired by Hamilton's Ricci flow~\cite{hamilton}, in
which $\tau$ is an auxiliary relaxation parameter. Since the axial
contorsion leaves the symmetric part of the Ricci tensor unchanged at
linear order (Section~\ref{sec:modified-flow}), Eq.~\eqref{eq:flow}
splits into
\begin{equation}\label{eq:flow-split}
\partial_\tau h_{(\mu\nu)}=-2R_{(\mu\nu)}(\mathring\Gamma),
\qquad
\partial_\tau h_{[\mu\nu]}=-2R_{[\mu\nu]}(\Gamma).
\end{equation}
The first equation is the linearized metric Ricci flow. The second is
driven by torsion and is the part considered in Refs.~\cite{r1,r2}
and in this work.

In Ref.~\cite{r1}, stationary configurations satisfy integrated
conditions: the chiral currents need not vanish locally, provided that
their antisymmetric contribution cancels globally. Such configurations
organize into toroidal domains dominated by one chirality, interpreted
as gravitational skyrmions~\cite{r1}. In Ref.~\cite{r2} these domains
are stabilized by a topological charge associated with $\pi_3(S^3)$.
Between domains of opposite chirality, the flip sector gives a
torsional vortex~\cite{r1} and a M\"obius-like bridge, in which a
$\mathbb Z_2$ twist acts on the internal spinorial frame while
spacetime remains orientable~\cite{r2}. Reference~\cite{r1} also
showed that the vector potential of toroidal currents can be nonzero
where the curvature vanishes, and that, under the stationarity
conditions of the flow, the flip term integrates to zero and gives no
static contribution, while oscillatory effects may remain.

\subsection{From the Dirac Lagrangian to torsion}

The spin connection enters the Dirac Lagrangian through the covariant
derivative, and its coupling to the spinor reads~\cite{r3}
\begin{equation}\label{eq:Dirac-coupling}
\mathcal L_\omega=-\tfrac18\,\omega_{\mu ab}J^{\mu ab},
\qquad
J^{\mu ab}=\bar\psi\{\gamma^\mu,\sigma^{ab}\}\psi .
\end{equation}
By the Clifford identity
$\{\gamma^\mu,\sigma^{ab}\}=\kappa_A\epsilon^{\mu ab\rho}\gamma_\rho\gamma^5$,
the source is totally antisymmetric and contains only the axial
current~\cite{ost2014},
\begin{equation}\label{eq:axial-source}
J^{\mu ab}=\kappa_A\epsilon^{\mu ab\rho}J_{A\rho},
\qquad
J_A^\rho=\bar\psi\gamma^\rho\gamma^5\psi .
\end{equation}
Its temporal component is the chiral imbalance $J_A^0=n_R-n_L$, its
spatial components are proportional to the current difference $J^k$
of Eq.~\eqref{eq:chiral-current-difference}, and for $m\neq0$ it
satisfies $\partial_\mu J_A^\mu=2im\,S$, with $S$ the mixed bilinear
of Eq.~\eqref{eq:mixed-chiral-bilinear}.

In the first-order formulation the spin connection is independent of
the tetrad~\cite{hehl,shapiro,hop2013},
\begin{equation}\label{eq:connection}
\omega_{\mu ab}=\mathring\omega_{\mu ab}(e)+K_{\mu ab},
\end{equation}
and the coupling~\eqref{eq:Dirac-coupling} splits into
$-\tfrac18\mathring\omega_{\mu ab}J^{\mu ab}-\tfrac18K_{[\mu ab]}J^{\mu ab}$,
since only the axial projection of the contorsion couples
to~\eqref{eq:axial-source}. The torsion is the antisymmetric part of
the connection, $T^\lambda{}_{\mu\nu}=K^\lambda{}_{\mu\nu}-K^\lambda{}_{\nu\mu}$,
and in the axial sector
\begin{equation}\label{eq:torsion-contorsion}
T_{\mu ab}=-2K_{[\mu ab]},
\qquad
\mathcal L_{\mathrm{int}}=\tfrac1{16}\,T_{\mu ab}J^{\mu ab}.
\end{equation}
The independent connection determines the torsion through the
contorsion. In the weak-field approximation, we adopt, as in Refs.~\cite{r1,r2,r3},
an effective identification between the antisymmetric tetrad
perturbation and the axial projection of torsion. The relation below
is understood through its totally antisymmetric projection, rather than
as an identification of the full torsion tensor.
\begin{equation}\label{eq:torsion-tetrad}
T_{\mu ab}=\tfrac12\bigl(\partial_b h_{[\mu a]}-\partial_a h_{[\mu b]}\bigr).
\end{equation}
Its totally antisymmetric part and the corresponding contorsion are
\begin{equation}\label{eq:axial-tetrad}
T_{[\rho\mu\nu]}=\partial_{[\rho}h_{\mu\nu]},
\qquad
K_{[\rho\mu\nu]}=-\tfrac12\,\partial_{[\rho}h_{\mu\nu]}.
\end{equation}

\subsection{The modified Ricci flow}\label{sec:modified-flow}

At linear order about flat space, the totally antisymmetric contorsion
contributes only to the antisymmetric part of the Ricci tensor,
\begin{equation}\label{eq:Ranti}
R_{(\mu\nu)}(\Gamma)=R_{\mu\nu}(\mathring\Gamma),\qquad
R_{[\mu\nu]}(\Gamma)=-\partial^\rho K_{\rho\mu\nu}.
\end{equation}
The torsional part of the flow~\eqref{eq:flow-split} is therefore
\begin{equation}\label{eq:torsional-flow}
\partial_\tau h_{[\mu\nu]}=2\,\partial^\rho K_{\rho\mu\nu}
=-\partial^\rho T_{\rho\mu\nu},
\end{equation}
and with Eq.~\eqref{eq:axial-tetrad}
\begin{equation}\label{eq:torsional-flow-h}
\partial_\tau h_{[\mu\nu]}
=-\tfrac13\bigl(\Box h_{[\mu\nu]}
+\partial_\mu\partial^\rho h_{[\nu\rho]}
+\partial_\nu\partial^\rho h_{[\rho\mu]}\bigr).
\end{equation}
For divergence-free perturbations in the static limit,
Eq.~\eqref{eq:torsional-flow-h} reduces to
$\partial_\tau h_{[\mu\nu]}=\tfrac13\nabla^2h_{[\mu\nu]}$, a heat
equation for the antisymmetric sector, analogous to the metric Ricci
flow in harmonic gauge.

The spinor source enters through the dynamics of the axial torsion,
given in Section~\ref{sec:yukawa},
$(\Box+M_T^2)\,T_{\mu ab}\propto J_{\mu ab}$. With
Eq.~\eqref{eq:axial-tetrad} this becomes
\begin{equation}\label{eq:source-h}
(\Box+M_T^2)\,\partial_{[\rho}h_{\mu\nu]}\propto J_{\rho\mu\nu},
\end{equation}
which has the structure of the equation for the tetrad derivative in
Ref.~\cite{r3}. Section~\ref{sec:published} integrates it to obtain
$h_{[\mu\nu]}$.

A fixed point of Eq.~\eqref{eq:torsional-flow} requires
$\partial^\rho T_{\rho\mu\nu}=0$. For the screened solution of
Section~\ref{sec:yukawa}, written in terms of the axial potential
$\mathcal A^\sigma$ of Eq.~\eqref{eq:Kstatic}, this reads
\begin{equation}\label{eq:fixed-point}
\epsilon_{\rho\mu\nu\sigma}\,\partial^\rho\mathcal A^\sigma=0,
\end{equation}
so the curl of the screened axial potential must vanish. This is the
local form of the stationarity conditions of Ref.~\cite{r1}. In the
unscreened limit it holds pointwise outside the sources. With
screening it fails within a layer of thickness of order $M_T^{-1}$
around them, and inside a source region it can hold only in the
integrated sense of Ref.~\cite{r1}.

\section{Spin sources and screened integral representations}\label{sec:yukawa}

The integral representations of Ref.~\cite{r2} use a Yukawa kernel to
describe the static response to a spinor source. We first show how
a screening scale arises in the effective axial torsion sector and
then specify the screened response used in the toroidal calculation.

For the axial torsion of Eq.~\eqref{eq:torsion-contorsion}, we use the effective
Ginzburg--Landau Lagrangian
\begin{equation}
\label{eq:LT}
\mathcal L_T
=\tfrac14\,\partial_\nu T_{\mu ab}\,\partial^\nu T^{\mu ab}
-\mu_{\mathrm{GL}}^2(T\!\cdot T)
-\lambda(T\!\cdot T)^2
+\tfrac1{16}T_{\mu ab}J^{\mu ab},
\qquad
T\!\cdot T=T_{\mu ab}T^{\mu ab}.
\end{equation}
Here $\mu_{\mathrm{GL}}^2$ is the quadratic coefficient of the
torsion potential and $\lambda$ is its quartic coupling.
Using the antisymmetry of $J_{\mu ab}$ in $a,b$, variation with
respect to $T_{\mu ab}$ gives
\begin{equation}
\Box T_{\mu ab}
+4\mu_{\mathrm{GL}}^2 T_{\mu ab}
+8\lambda(T\!\cdot T)T_{\mu ab}
=\tfrac18 J_{\mu ab}.
\end{equation}

For $\mu_{\mathrm{GL}}^2=-m_T^2$, a homogeneous nonzero condensate
$T^{(0)}$ in the absence of a source satisfies
$T^{(0)}\!\cdot T^{(0)}=m_T^2/(2\lambda)$.
Writing $T=T^{(0)}+\delta T$ and keeping terms linear in $\delta T$
gives
\begin{equation}\label{eq:GLlinear}
\Box\,\delta T_{\mu ab}
+16\lambda\bigl(T^{(0)}\!\cdot\delta T\bigr)T^{(0)}_{\mu ab}
=\tfrac18 J_{\mu ab}.
\end{equation}
The term proportional to $\lambda$ acts on the fluctuation along
$T^{(0)}$. In that direction the mass squared is
\[
M_T^2=16\lambda\bigl(T^{(0)}\!\cdot T^{(0)}\bigr)=8m_T^2.
\]
Fluctuations orthogonal to $T^{(0)}$ receive zero mass contribution
from this term. The parameters $m_T$ and $M_T$ refer to the torsion
sector, and $m$ denotes the fermion mass.

In the static limit, the Green function of the massive channel is
\begin{equation}\label{eq:GM}
(-\nabla^2+M_T^2)G_{M_T}(\bx)=\delta^{(3)}(\bx),
\qquad
G_{M_T}(r)=\frac{e^{-M_Tr}}{4\pi r}.
\end{equation}
For a localized source and decay at infinity, the fluctuation in
this channel is the convolution of $G_{M_T}$ with the projection
of $J/8$ along $T^{(0)}$. The corresponding axial contorsion
fluctuation is $\delta K=-\delta T/2$.

To compare with Ref.~\cite{r2}, we match its inverse coherence length
$\mu$ to the screening scale $M_T$. For the comparison with the
published integrals and for the toroidal calculation, we use the
componentwise ansatz
\begin{equation}\label{eq:Kstatic}
\mathcal K_{\lambda\mu\nu}(\bx)
=\epsilon_{\lambda\mu\nu\sigma}\,\mathcal A^\sigma(\bx),
\qquad
\mathcal A^\sigma(\bx)
=\kappa_A\int G_{M_T}(\bx-\bx')J_A^\sigma(\bx')\,d^3x'.
\end{equation}
Here $\mathcal A^\sigma$ is the screened axial potential and
$\kappa_A$ sets the normalization of the effective response.
The linearized equation \eqref{eq:GLlinear} gives the screening
mass $M_T$ only for the fluctuation along the condensate. Applying the
same kernel to every component of $J_A^\sigma$ is an effective
assumption of this ansatz. The Lagrangian~\eqref{eq:LT} is used as an
effective model and does not provide a complete analysis of the
dynamical consistency of the torsion field.

Within this ansatz, Eq.~\eqref{eq:Kstatic} is the screened solution
of Eq.~\eqref{eq:source-h}. The fixed-point
condition~\eqref{eq:fixed-point} is a condition on the curl of
$\mathcal A^\sigma$. The spatial part of $\mathcal A^\sigma$ is the
vector potential of Section~\ref{sec:torus}, and its temporal part is
the potential $A^0$ of Section~\ref{sec:fluct}.

In terms of $\mathcal A^\sigma$, the interaction term of
Eq.~\eqref{eq:torsion-contorsion} takes a simple form. With
Eqs.~\eqref{eq:axial-source} and~\eqref{eq:Kstatic},
$T_{\mu ab}=-2\epsilon_{\mu ab\sigma}\mathcal A^\sigma$ and
$\epsilon_{\mu ab\sigma}\epsilon^{\mu ab\rho}=-6\,\delta_\sigma^\rho$, so that
\begin{equation}\label{eq:Lint-A}
\mathcal L_{\mathrm{int}}=\tfrac1{16}\,T_{\mu ab}J^{\mu ab}
=\tfrac34\,\kappa_A\,\mathcal A_\mu\,\bar\psi\gamma^\mu\gamma^5\psi ,
\end{equation}
up to the sign convention for $\epsilon_{0123}$. The fermion couples to
$\mathcal A_\mu$ as a charged particle couples to an electromagnetic
potential, with $\gamma^\mu\gamma^5$ in place of $\gamma^\mu$. The same
potential therefore determines the tetrad response and enters the
spinor dynamics of Section~\ref{sec:fluct}.

Section~\ref{sec:published} compares Eq.~\eqref{eq:Kstatic} with the
published integral representations and discusses the conditions under
which the additional line integration reconstructs the tetrad
perturbation.
\section{Integral representations}\label{sec:published}

The stationary integral expressions in Ref.~\cite{r2}, Eqs.~(20)--(22),
use the screened Green kernel discussed in Section~\ref{sec:yukawa}.
We develop the three integral representations starting from the
weak-field Lagrangian formulation of Ref.~\cite{r3}. In that formulation,
the Green function gives the response of a tetrad derivative, which
is then integrated to obtain the corresponding tetrad perturbation.

In the linearized spinor Lagrangian, the spin-connection term couples
the tetrad derivative
\begin{equation}\label{eq:tetrad-derivative}
h_{\mu c|b}=\partial_b h_{\mu c}
\end{equation}
to the spinor bilinear $\bar\psi\{\gamma^\mu,\sigma^{bc}\}\psi$,
with the index ordering and normalization of Ref.~\cite{r3}.
The anticommutator gives the axial Levi-Civita structure discussed
in Section~\ref{sec:flow}. The field equation for $h_{\mu c|b}$
and its Green-function solution are given in Eqs.~(56)--(62) of
Ref.~\cite{r3}. Retaining the sourced particular solution, we write
\begin{equation}\label{eq:tetrad-derivative-green}
(\partial_b h_{\mu c})_{\mathrm{part}}(x)
=\int d^4x'\,G(x,x')\,S_{\mu c|b}(x'),
\end{equation}
where $S_{\mu c|b}$ is the projected spinor source in the
normalization of Ref.~\cite{r3}. Here $G(x,x')$ is the space-time Green
function of that representation. The stationary expressions in
Ref.~\cite{r2} use the spatial Yukawa kernel of Section~\ref{sec:yukawa}.

The source projections in Ref.~\cite{r3}, Eqs.~(68)--(76), separate
the left-handed, right-handed, and mixed contributions. The first
two contributions enter through the current difference $J^k$, and
the third through the mixed chiral bilinear $S$,
\begin{align}
J^k
&=\bar\psi_L\sigma^k\psi_L-\bar\psi_R\sigma^k\psi_R,
\label{eq:chiral-current-difference}\\
S
&=\bar\psi_L\psi_R-\bar\psi_R\psi_L.
\label{eq:mixed-chiral-bilinear}
\end{align}
With the derivative index after the vertical bar, the three
projections used in Eq.~\eqref{eq:tetrad-derivative-green} are
\begin{align}
S_{0c|b}
&=\tfrac13\,\epsilon_{bck}J^k,
\label{eq:source-mixed-tetrad}\\
S^{(L/R)}_{ac|0}
&=-\tfrac13\,\epsilon_{ack}J^k,
\label{eq:source-spatial-current}\\
S^{(\mathrm{flip})}_{ac|b}
&=-\tfrac13\,\epsilon_{bca}S.
\label{eq:source-spatial-flip}
\end{align}

For the first and third projections, we integrate the spatial
derivative data along a path at fixed time, contracting $b$ with
$dx^b$. For the second projection, we integrate in $x^0$ at
fixed spatial position. Setting the reference values of the
sourced contributions to zero gives the three representations
corresponding to Eqs.~(83)--(85) of Ref.~\cite{r3},
\begin{align}
h_{[0c]}(x)
&=\frac13\int dx^b
  \Big[\int d^4x'\,G(x,x')\,\epsilon_{bck}J^k(x')\Big],
\label{eq:integral-h0c}\\
h_{[ac]}(x)
&=-\frac13\int dx^0
  \Big[\int d^4x'\,G(x,x')\,\epsilon_{ack}J^k(x')\Big],
\label{eq:integral-hac-current}\\
h_{[ac]}(x)
&=-\frac13\int dx^b
  \Big[\int d^4x'\,G(x,x')\,\epsilon_{bca}S(x')\Big].
\label{eq:integral-hac-flip}
\end{align}
In each expression, the unprimed argument of $G$ is evaluated at
the point of the outer integration. Equation~\eqref{eq:integral-hac-flip} coincides with Eq.~(85) of
Ref.~\cite{r3}, and we denote this flip contribution by
$h^{(\mathrm{flip})}_{[ac]}$. Equations~\eqref{eq:integral-hac-current}
and~\eqref{eq:integral-hac-flip} give separate contributions to $h_{[ac]}$.

These sourced contributions belong to the antisymmetric tetrad
sector. In Eqs.~\eqref{eq:integral-hac-current}
and~\eqref{eq:integral-hac-flip}, exchanging $a$ and $c$ reverses
the sign of the Levi-Civita factor. For Eq.~\eqref{eq:integral-h0c},
the axial derivative data with $0$ and $c$ exchanged have the opposite
sign at fixed $b$. With the same Green kernel, integration
prescription, and zero reference values, each sourced contribution is
therefore antisymmetric in its two indices. This is why we write them
as $h_{[0c]}$ and $h_{[ac]}$.

For Eqs.~\eqref{eq:integral-h0c} and~\eqref{eq:integral-hac-flip},
path-independent reconstruction requires the corresponding spatial
one-forms to be exact. Equation~\eqref{eq:integral-hac-current}
defines a temporal primitive once its reference value is specified.

The two current integrals, Eqs.~\eqref{eq:integral-h0c}
and~\eqref{eq:integral-hac-current}, describe how the chiral current
difference contributes to the mixed and spatial tetrad components.
The flip integral, Eq.~\eqref{eq:integral-hac-flip}, depends on the
overlap and relative phase of the chiral components.
Section~\ref{sec:torus} examines the screened response on the
stationary toroidal configuration of Ref.~\cite{r1}. Section~\ref{sec:fluct} examines the role of the fermion mass and of
helicity conversion, and the conditions under which the mixed source
$S$ becomes nonzero.
\section{Toroidal sources and torsional circulation}\label{sec:torus}

\subsection{The vector potential of the current integrals}

For a stationary source, the time integration in the inner convolution
of Eqs.~\eqref{eq:integral-h0c} and~\eqref{eq:integral-hac-current}
gives the static Yukawa kernel of Eq.~\eqref{eq:GM}. Both integrals
then contain the same vector potential,
\begin{equation}\label{eq:vector-potential}
A^k(\bx)=\int G_{M_T}(\bx-\bx')\,J^k(\bx')\,d^3x',
\end{equation}
with $J^k$ the current difference of
Eq.~\eqref{eq:chiral-current-difference}. In the unscreened limit
$M_T\to0$, $A^k$ is the magnetostatic vector potential of $J^k$.
Since $J_A^k$ is proportional to $J^k$, $A^k$ is proportional to the
spatial part of the axial potential $\mathcal A^\sigma$ of
Eq.~\eqref{eq:Kstatic}.
The inner brackets of the two integrals give
\begin{equation}\label{eq:potential-derivatives}
\partial_b h_{[0c]}=\tfrac13\,\epsilon_{bck}A^k,
\qquad
\partial_0 h_{[ac]}=-\tfrac13\,\epsilon_{ack}A^k .
\end{equation}
In Eq.~\eqref{eq:integral-h0c} the potential is integrated along a
spatial path and gives the mixed component $h_{[0c]}$. In
Eq.~\eqref{eq:integral-hac-current} it is integrated in time and
gives the spatial component $h_{[ac]}$. With zero reference value at
$t=0$,
\begin{equation}\label{eq:triad-rotation}
h_{[ac]}(t,\bx)=-\frac t3\,\epsilon_{ack}A^k(\bx).
\end{equation}
Here stationary refers to the flow parameter $\tau$ of
Eq.~\eqref{eq:flow}: the configuration is a fixed point of the flow.
In physical time, Eq.~\eqref{eq:triad-rotation} describes a rotation
of the spatial triad about the direction of $A^k$ at a constant rate
proportional to $|A|$.
The complete set of spatial one-forms in Eq.~\eqref{eq:integral-h0c}
is exact only for uniform $A^k$, so for
a nonuniform potential $h_{[0c]}$ is not reconstructed globally; the
results below use $A^k$, its circulation, and the temporal
reconstruction \eqref{eq:triad-rotation}, which holds pointwise.
The linear growth in physical time described by
Eq.~\eqref{eq:triad-rotation} is valid only within the weak-field
regime, where the accumulated tetrad perturbation remains sufficiently
small for the linear approximation to hold.

\subsection{Gravitomagnetic comparison and poloidal source}

In weak-field gravitomagnetism, a mass current with nonzero angular
momentum produces a dipolar contribution to the Lense--Thirring
field~\cite{lt}. The comparison introduced in Ref.~\cite{r2} motivates a
multipole analysis of the field sourced by chiral spin currents.
The stationary configurations of the Ricci flow obtained in
Ref.~\cite{r1} are toroidal domains, and we model each domain by a
poloidal current on a finite torus. For this source,
the magnetic-type dipole vanishes and the toroidal
moment~\cite{dubovik} is nonzero.
We calculate the circulation of the resulting axial field.

Let $\mathcal S$ be the surface of a torus of major radius $R$ and
minor radius $a<R$. In cylindrical coordinates $(\rho,\phi,z)$,
the meridional section is parametrized by
\begin{equation}\label{eq:torus-coordinates}
\rho=R+a\cos\theta,\qquad z=a\sin\theta.
\end{equation}
Following Ref.~\cite[Eq.~(25)]{r2}, we take the surface current to be
\begin{equation}\label{eq:torus-current}
\mathbf K=\frac{K_0}{\rho}\,\mathbf e_\theta,
\end{equation}
where $\mathbf e_\theta$ points in the direction of increasing
$\theta$. This current models the spatial profile of $J_L-J_R$.
Replacing a purely left-handed current with a purely right-handed
current of the same profile and direction sends $K_0$ to $-K_0$.
\begin{proposition}\label{prop:moments}
The magnetic-type dipole moment vanishes,
$\mathbf m=\tfrac12\int_{\mathcal S}\bx\times\mathbf K\,dS=0$.
The toroidal moment is
\begin{equation}\label{eq:T}
\bT=\frac1{10}\int_{\mathcal S}
\bigl[(\bx\cdot\mathbf K)\bx-2|\bx|^2\mathbf K\bigr]\,dS
=-\pi^2 K_0 R a^2\,\hat{\mathbf z}.
\end{equation}
\end{proposition}

\begin{proof}
The $z$-component of $\bx\times\mathbf K$ vanishes pointwise,
and the transverse components integrate to zero in $\phi$.
By axial symmetry, only the $z$-component of $\bT$ survives.
Using $dS=a\rho\,d\theta\,d\phi$,
$\bx\cdot\mathbf e_\theta=-R\sin\theta$, and
$(\mathbf e_\theta)_z=\cos\theta$, we obtain
\begin{equation}
\bT\cdot\hat{\mathbf z}
=\frac{2\pi K_0a}{10}
\int_0^{2\pi}
\bigl[-Ra\sin^2\theta
-2(R^2+a^2+2Ra\cos\theta)\cos\theta\bigr]\,d\theta
=-\pi^2K_0Ra^2 .
\end{equation}
\end{proof}

\subsection{Circulation with and without Yukawa screening}

Using the componentwise axial response of
Section~\ref{sec:yukawa}, we define
\begin{align}
\bA_{M_T}(\bx)
&=\int_{\mathcal S}G_{M_T}(\bx-\bx')\,\mathbf K(\bx')\,dS',
\label{eq:torus-axial-field}\\
\mathbf B_{M_T}
&=\nabla\times\bA_{M_T},
\label{eq:torus-curl}\\
\Phi_{\mathcal C}(M_T)
&=\oint_{\mathcal C}\bA_{M_T}\cdot d\bx.
\label{eq:torus-circulation}
\end{align}
Here $G_{M_T}$ is the Yukawa kernel of Eq.~\eqref{eq:GM}, the
limit $M_T\to0$ gives the unscreened response, and the overall
normalization is absorbed into $K_0$. Equation~\eqref{eq:torus-axial-field}
is the vector potential of Eq.~\eqref{eq:vector-potential} for the torus
current. By axial symmetry it lies in the meridional planes. The contour $\mathcal C$ lies outside the solid torus.
By Eq.~\eqref{eq:Ranti}, the antisymmetric Ricci component
$R_{[0c]}$ is proportional to $(\mathbf B_{M_T})_c$.
Linearity gives $\Phi_{\mathcal C}\mapsto-\Phi_{\mathcal C}$ under
the exchange of the left- and right-handed sources described above,
with the contour orientation held fixed.

\begin{proposition}\label{prop:circ}
Let $N$ be the signed linking number of $\mathcal C$ with the core
circle of the torus. Choose its positive sense so that a single
positive linking gives positive circulation for $K_0>0$.
In the unscreened limit,
\begin{equation}\label{eq:Phi}
\Phi_{\mathcal C}(0)
=2\pi N K_0\bigl(R-\sqrt{R^2-a^2}\bigr).
\end{equation}
For $a/R\ll1$, the leading term is
\begin{equation}\label{eq:Phi-thin}
\Phi_{\mathcal C}(0)\simeq
-\frac{N\,\bT\cdot\hat{\mathbf z}}{\pi R^2}.
\end{equation}
With Yukawa screening, contours with the same linking number
can have different circulations.
\end{proposition}

\begin{proof}
For $M_T=0$, Amp\`ere's law for the prescribed poloidal current
gives $\mathbf B_0=-K_0\mathbf e_\phi/\rho$ inside the tube and
$\mathbf B_0=0$ outside. The positive normal to a meridional disk
is $-\mathbf e_\phi$. By Stokes' theorem, a single positive linking
gives the flux through that disk,
\begin{equation}\label{eq:torus-meridional-flux}
K_0\int_{-a}^{a}
\ln\!\left(\frac{R+\sqrt{a^2-z^2}}{R-\sqrt{a^2-z^2}}\right)\,dz
=2\pi K_0\bigl(R-\sqrt{R^2-a^2}\bigr).
\end{equation}
The linking number gives the signed multiplicity of this flux.
Expansion in $a/R$ and substitution of Eq.~\eqref{eq:T}
give Eq.~\eqref{eq:Phi-thin}.

For $M_T>0$, the unscreened and screened Green functions give,
for the same source,
\begin{equation}\label{eq:torus-screened-curl}
\mathbf B_{M_T}=\mathbf B_0-M_T^2G_{M_T}*\mathbf B_0.
\end{equation}
The convolution extends the azimuthal response outside the tube.
At an exterior point off the symmetry axis, pair source elements
at the same $\rho'$ and $z'$ whose azimuthal directions have
opposite projections onto the local $\mathbf e_\phi$. The closer
element receives greater weight from $G_{M_T}$, so the paired
contributions fail to cancel for $K_0\ne0$. By Stokes' theorem,
a deformation of the contour through this exterior field can
therefore change the circulation at fixed $N$.
\end{proof}

Equation~\eqref{eq:Phi} gives the circulation for a finite torus.
Equation~\eqref{eq:Phi-thin} connects it to the toroidal moment
in the thin-torus limit, addressing the relation left open in
Ref.~\cite{r2}. In the unscreened case, an exterior contour can have
nonzero circulation while the curl vanishes along it. This is
the basis of the Aharonov--Bohm analogy~\cite{ab}. A phase for a transported
spinor requires a separate calculation of its coupling to the
axial field.

\paragraph{Stationary domains on tori.}
In the stationary configurations of Refs.~\cite{r1,r2}, left- and
right-handed domains occupy separate tori. Reversing the chirality
reverses $K_0$, so the two domains have opposite potentials and
opposite circulations. By Eq.~\eqref{eq:triad-rotation}, the spatial
triad rotates in opposite senses in the two domains. In the unscreened
case, the circulation of the rotation rate along a contour that
links one tube is
$\pm\Phi_{\mathcal C}(0)/3$, and it is present also outside the
tube, where the antisymmetric Ricci component vanishes. By
Eq.~\eqref{eq:fixed-point}, this exterior region satisfies the
fixed-point condition pointwise: the stationary configuration carries
a nonzero vector potential where the curvature vanishes, as anticipated
in Ref.~\cite{r1}. For two
coaxial tori with opposite currents placed symmetrically about the
plane $z=0$, the $z$-components of the two potentials cancel on that
plane, while the radial components add. On the symmetry axis the
potential therefore vanishes at $z=0$.

Screening changes this picture in three ways. First, each potential
decays outside its tube on the scale $M_T^{-1}$, so that domains
separated by more than this distance act independently. Second, the
antisymmetric Ricci component is no longer confined to the tube and
extends outside it over a layer of thickness of order $M_T^{-1}$, by the same
mechanism that limits a magnetic field in a superconductor to the
London depth. Third, the circulation, and hence the integrated
rotation of the triad, depends on the contour and decreases with its
distance from the tube, whereas without screening it takes the value
of Eq.~\eqref{eq:Phi} on every contour that links the tube once. This
decrease carries the screening scale $M_T$.

Section~\ref{sec:fluct} turns to the mixed source $S$ in
Eq.~\eqref{eq:integral-hac-flip} and examines the conditions under
which it can provide a source at an interface between domains.

\section{Chiral mixing and the antisymmetric flip response}
\label{sec:fluct}

The poloidal spin current considered in Section~\ref{sec:torus}
generates a spatial axial field in the meridional plane. For a
fermion moving in the azimuthal direction, around the major circle
of the torus, this spatial field is transverse to the momentum.
This geometry motivates the study of helicity conversion in a
transverse axial background. We first examine the mechanism in a
uniform field. We then evaluate the flip integral
\eqref{eq:integral-hac-flip} for a prescribed mixed-chirality
source at an interface.

\subsection{Mass coupling and chiral weights}

\begin{figure}[htbp]
\centering
\includegraphics[width=0.42\textwidth]{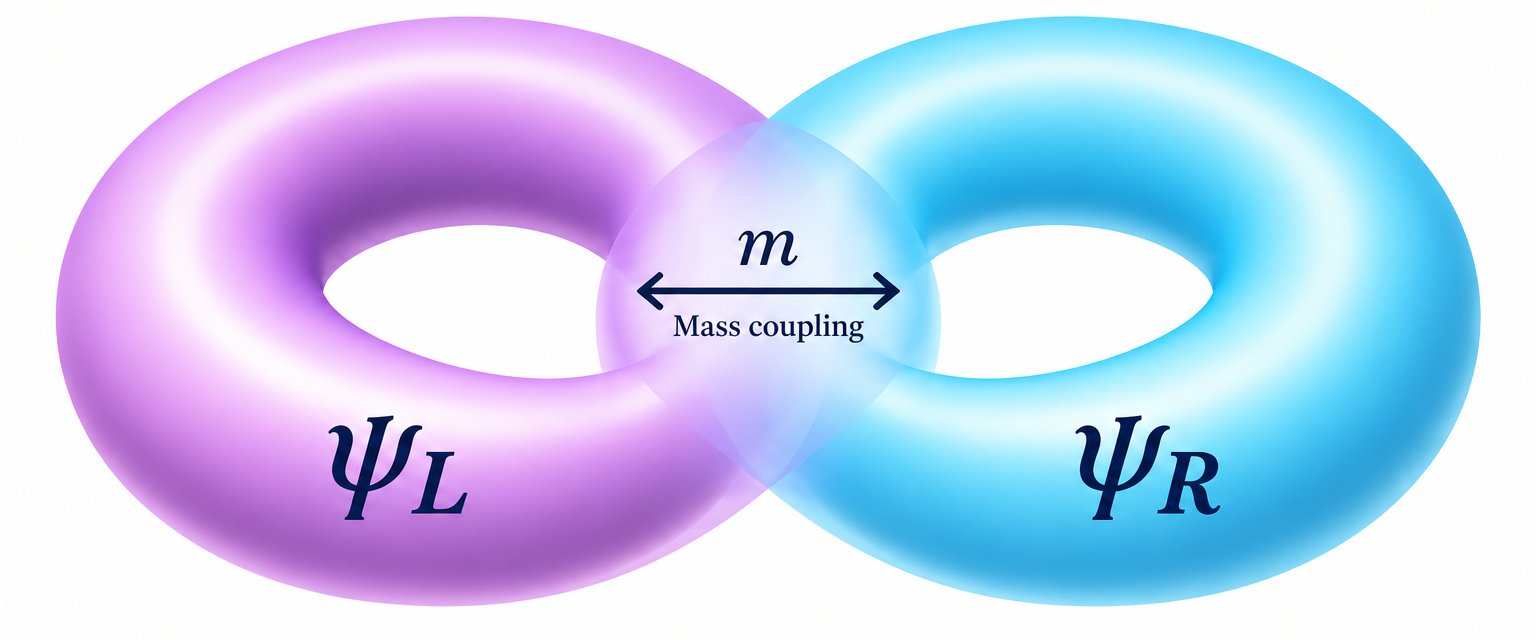}
\caption{The mass term $m$ couples the left- and right-handed
components of a single spinor.}
\label{fig:mass}
\end{figure}

For the helicity branch considered here, a uniform, purely temporal
axial background gives
\begin{equation}\label{eq:twolevel}
i\frac{d}{dt}
\begin{pmatrix}\psi_L\\ \psi_R\end{pmatrix}
=
\begin{pmatrix}-\Delta&m\\m&\Delta\end{pmatrix}
\begin{pmatrix}\psi_L\\ \psi_R\end{pmatrix},
\qquad
\Delta=p-A^0,\qquad p=|\mathbf p|,
\end{equation}
in units with $\hbar=c=1$. The mass $m$ couples the two chiral
amplitudes, as illustrated in Fig.~\ref{fig:mass}. The temporal
axial potential shifts their diagonal energy difference, as in the
constant axial background of Ref.~\cite{ck1997}. Helicity is
conserved in this background.

\begin{figure}[htbp]
\centering
\includegraphics[width=0.5\textwidth]{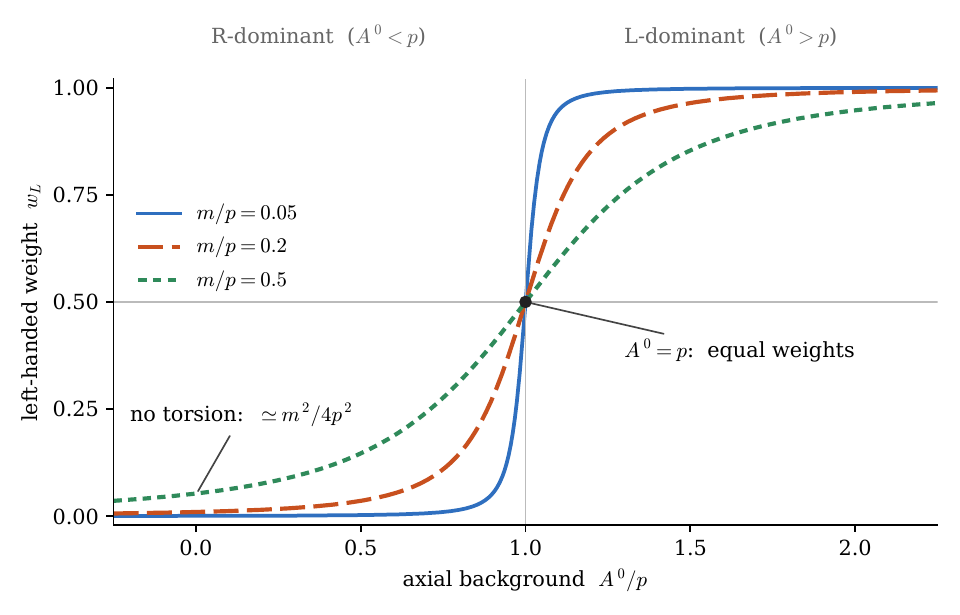}
\caption{Left-handed weight of the positive-energy eigenstate
at fixed helicity as a function of $A^0/p$, for three values of
$m/p$. Each point represents a uniform axial background.
The weights are equal at $A^0=p$. The mixed source $S$ vanishes
for the eigenstates shown: equal chiral weights do not imply a flip
source.}
\label{fig:scales}
\end{figure}

For the positive-energy eigenstate, with
$\Omega=\sqrt{m^2+\Delta^2}$, the normalized chiral weights are
\begin{equation}\label{eq:mix}
w_L=\frac12\left(1-\frac{\Delta}{\Omega}\right),\qquad
w_R=\frac12\left(1+\frac{\Delta}{\Omega}\right).
\end{equation}
For $m>0$, the two weights are equal at $A^0=p$. The mixed
bilinear $S$ of Eq.~\eqref{eq:mixed-chiral-bilinear} is purely
imaginary, $S=2i\,\operatorname{Im}(\psi_L^\dagger\psi_R)$.
In the eigenstate of Eq.~\eqref{eq:twolevel}, the two components
share the same helicity spinor and their relative coefficient can
be chosen real. Hence $S=0$ for the eigenstates shown in
Fig.~\ref{fig:scales}, including the state with equal chiral weights.

\subsection{Helicity conversion and the mixed source}

In the rest frame of the source, the Hamiltonian in a uniform
axial field $(A^0,\mathbf A)$ is
\begin{equation}\label{eq:fourlevel}
H=
\begin{pmatrix}
-\boldsymbol{\sigma}\cdot(\mathbf p-\mathbf A)+A^0 I_2 & mI_2\\
mI_2 & \boldsymbol{\sigma}\cdot(\mathbf p+\mathbf A)-A^0 I_2
\end{pmatrix}.
\end{equation}
For $p>0$, we decompose the spatial field relative to the momentum,
\begin{equation}\label{eq:axial-field-decomposition}
A_\parallel=\hat{\mathbf p}\cdot\mathbf A,\qquad
\mathbf A_\perp=\mathbf A-A_\parallel\hat{\mathbf p},\qquad
A_\perp=|\mathbf A_\perp|.
\end{equation}
The transverse term
$\boldsymbol{\sigma}\cdot\mathbf A_\perp$ acts within each
chiral block and couples opposite spin projections along
$\hat{\mathbf p}$. At weak field, projection onto the two free
positive-energy helicity states gives an off-diagonal coupling
of magnitude $(m/E)A_\perp$, where $E=\sqrt{p^2+m^2}$. The
diagonal splitting is $2[A_\parallel-(p/E)A^0]$ at the same order.

For a purely transverse field, $A^0=A_\parallel=0$, the helicity
conversion probability for an initially definite-helicity state is
\begin{equation}\label{eq:flipfreq}
P_{\mathrm{flip}}(t)\simeq\sin^2(\omega t),\qquad
\omega\simeq\frac{m}{E}A_\perp.
\end{equation}
Torsion-induced helicity transitions have been studied in
Ref.~\cite{cils1999}; see also Ref.~\cite{dsg1981} for neutrino
oscillations in the presence of torsion. For an ultrarelativistic Majorana neutrino,
the opposite-helicity state can participate in weak interactions
as an antineutrino. Applying Eq.~\eqref{eq:flipfreq} along a
toroidal trajectory would require following the spinor in the
spatially varying field.

\paragraph{Source generated during the evolution.}
Let $n_L=\psi_L^\dagger\psi_L$ and $n_R=\psi_R^\dagger\psi_R$ be the
densities of the left- and right-handed components of the same state.
For a massive fermion these are not two separate particles: each state
contributes to both, and for a unit-normalized state at fixed momentum
$n_L$ and $n_R$ are the chiral weights, with $n_L+n_R=1$.
The diagonal blocks of
Eq.~\eqref{eq:fourlevel} preserve each chiral norm, while the
mass term gives
\begin{equation}\label{eq:flip-source-imbalance}
\frac{d}{dt}(n_R-n_L)=2im\,S.
\end{equation}
For a unit-normalized state with initial helicity $s=\pm1$,
the changing helicity weights give, to leading order in the
purely transverse case,
$n_R-n_L\simeq s(p/E)\cos(2\omega t)$. It follows that, for $m>0$,
\begin{equation}\label{eq:transverse-flip-source}
iS(t)\simeq-s\,\frac{A_\perp p}{E^2}\sin(2\omega t).
\end{equation}
The source oscillates with zero period average. Opposite initial
helicities give opposite sources in this approximation. For an
incoherent mixture with matched momentum, spatial profile, and
preparation time, a fraction $f$ of the opposite initial helicity
gives the net source $(1-2f)S(t)$ at fixed total density and fixed
background. Fluctuations in $f$ can therefore change the sign and
magnitude of the source entering Eq.~\eqref{eq:integral-hac-flip}.

\subsection{Screened response at a thin interface}\label{sec:bridge}

\begin{figure}[htbp]
\centering
\includegraphics[width=0.85\textwidth]{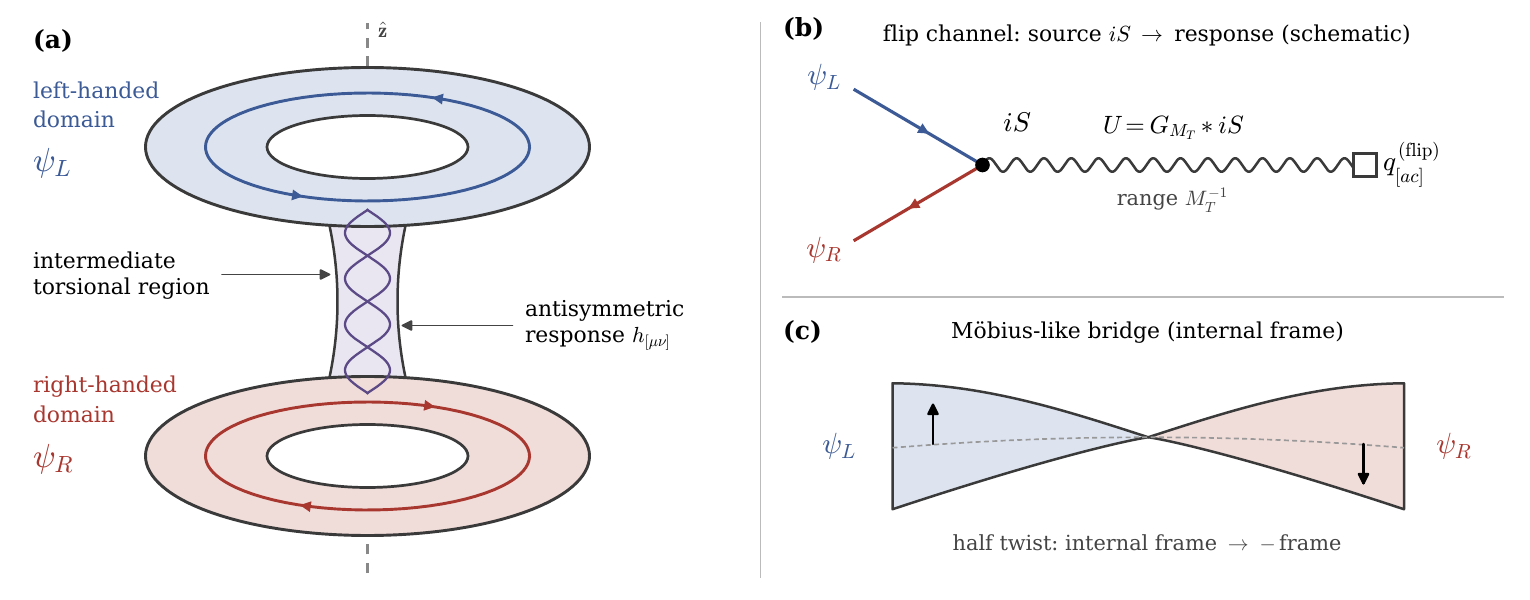}
\caption{Proposed interface between domains of opposite chiral
dominance. The panels show the intermediate torsional region,
a schematic view of the source--response relation in
Eq.~\eqref{eq:integral-hac-flip}, and a possible M\"obius-like
twist of the internal spinorial frame.}
\label{fig:bridge}
\end{figure}

Figure~\ref{fig:bridge} illustrates the interface between domains of
opposite chiral dominance proposed in Refs.~\cite{r1,r2}, where it was
interpreted as a M\"obius-like bridge of the internal spinorial frame.
Reference~\cite{r1} also noted that, under the stationarity
conditions of the flow, the flip term gives no static contribution,
and left its oscillatory effects for further study. To evaluate the flip
integral explicitly, we take a prescribed source localized at a planar interface at $z=0$,
static or oscillating slowly on the scale $M_T^{-1}$. The strength of this
source is an input to the calculation.

\begin{figure}[tb]
\centering
\includegraphics[width=0.85\textwidth]{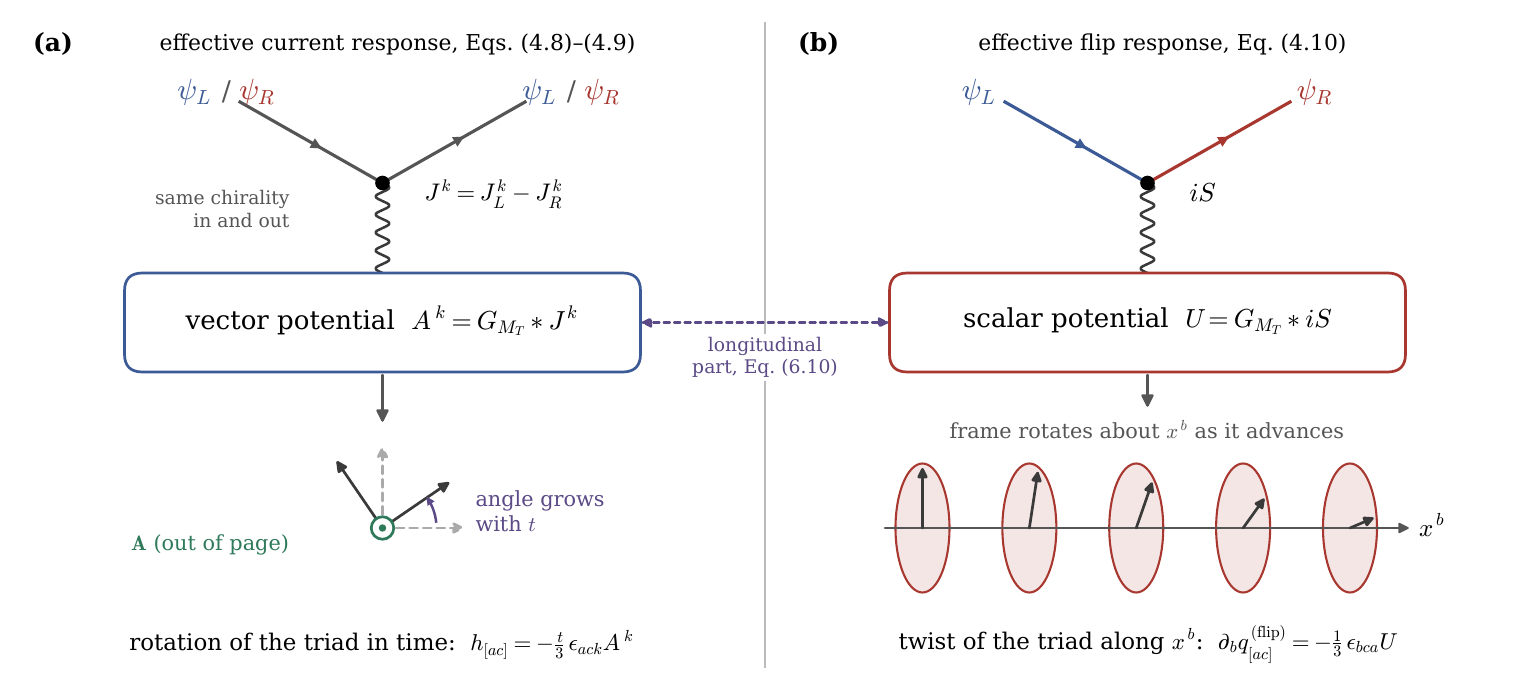}
\caption{Schematic comparison of the effective current and flip
responses. (a) The current difference sources the vector potential
$A^k$, which rotates the spatial triad in time, Eq.~\eqref{eq:triad-rotation}.
(b) The mixed bilinear sources the scalar potential $U$, which twists
the triad along one direction, Eq.~\eqref{eq:flip-scalar-potential}. The wavy lines represent the Green kernel
$G_{M_T}$, not a quantum propagator. $U$ is related to the longitudinal
part of the axial potential by Eq.~\eqref{eq:flip-drives-A0}, which
does not enter the rotation of panel (a).}
\label{fig:channels}
\end{figure}

\paragraph{Scalar potential of the flip integral.}
For a source static or slowly varying on the scale $M_T^{-1}$, the
inner convolution of
Eq.~\eqref{eq:integral-hac-flip} defines a scalar potential, in
contrast with the vector potential of the current integrals in
Section~\ref{sec:torus}. Since $S$ is imaginary, we introduce the
real source $iS$ and define
$q_{[ac]}^{(\mathrm{flip})}=i\,h_{[ac]}^{(\mathrm{flip})}$ as the real
response variable associated with the flip channel,
\begin{equation}\label{eq:flip-scalar-potential}
U(\bx)=\int G_{M_T}(\bx-\bx')\,iS(\bx')\,d^3x',
\qquad
\partial_b q_{[ac]}^{(\mathrm{flip})}
=-\tfrac13\,\epsilon_{bca}U .
\end{equation}
For fixed $a$ and $c$, the Levi-Civita factor selects the single
direction $b$ orthogonal to both. The variable
$q_{[ac]}^{(\mathrm{flip})}$ is an infinitesimal rotation of the
spatial triad about the $x^b$ axis, and Eq.~\eqref{eq:flip-scalar-potential}
states that this rotation accumulates as $\int U\,dx^b$ along that
axis. The scalar potential therefore twists the triad along one
direction. Path independence of a given component requires $U$ to
depend only on the corresponding coordinate $x^b$. Figure~\ref{fig:channels} compares this response
with the rotation produced by the vector potential.

\paragraph{Axial potentials and the flip source.}
The axial background in Eq.~\eqref{eq:fourlevel} is the dual of the
totally antisymmetric contorsion of Eq.~\eqref{eq:Kstatic}. For a
source that varies slowly on the scale $M_T^{-1}$, its components are
those of the axial potential $\mathcal A^\sigma$ of
Eq.~\eqref{eq:Kstatic},
\begin{equation}\label{eq:axial-potentials}
A^0(\bx)=\kappa_A\int G_{M_T}(\bx-\bx')\,(n_R-n_L)(\bx')\,d^3x',
\qquad
A^k_{\mathrm{ax}}(\bx)=\kappa_A\int G_{M_T}(\bx-\bx')\,J_A^k(\bx')\,d^3x'.
\end{equation}
Here $n_L$ and $n_R$ are the densities of the left- and right-handed
components of the same field, defined in
Eq.~\eqref{eq:flip-source-imbalance}. Their difference is the chiral
imbalance $J_A^0=\psi^\dagger\gamma^5\psi$. The spatial part $J_A^k$
is proportional to the current difference $J^k$ of
Eq.~\eqref{eq:chiral-current-difference}, so that
$A^k_{\mathrm{ax}}$ is proportional to the vector potential of
Eq.~\eqref{eq:vector-potential}. The components $A_\parallel$ and
$A_\perp$ of Eq.~\eqref{eq:axial-field-decomposition} are its
projections on the momentum. For a fermion moving along
$\mathbf e_\phi$ around the torus of Section~\ref{sec:torus}, the
potential lies in the meridional planes, so that $A_\parallel=0$ and
$A_\perp\propto|A|$.

As noted in Section~\ref{sec:flow}, the axial current is not
conserved for $m\neq0$,
$\partial_\mu(\bar\psi\gamma^\mu\gamma^5\psi)=2im\,\bar\psi\gamma^5\psi
=2im\,S$, and for a uniform state this reduces to
Eq.~\eqref{eq:flip-source-imbalance}. Convolving with the static
kernel gives
\begin{equation}\label{eq:flip-drives-A0}
\partial_tA^0+\partial_kA^k_{\mathrm{ax}}=2m\,\kappa_A\,U .
\end{equation}
The mixed source therefore drives the temporal axial potential. In a
static configuration, $\partial_tA^0=0$, and the flip potential is
fixed by the divergence of the spatial potential. For the poloidal
torus, $J^k$ is divergence-free, so $\partial_kA^k_{\mathrm{ax}}=0$ and
$U=0$: a static toroidal domain carries no flip source, in agreement
with Ref.~\cite{r1} and with $S=0$ for the eigenstates of
Section~\ref{sec:fluct}. A configuration that is stationary in the
sense of the flow need not be static. If the chiral imbalance
oscillates, $\partial_tA^0\neq0$ drives an oscillating flip potential,
as for the source of Eq.~\eqref{eq:transverse-flip-source}.

Within the identification~\eqref{eq:torsion-tetrad}, the axial potential
is the dual of an exact form and is divergence-free. Equation~\eqref{eq:flip-drives-A0}
therefore concerns the longitudinal part of $\mathcal A^\sigma$, sourced by
the mixed bilinear, which is not reconstructed from the tetrad and belongs
to the independent axial torsion of the first-order formulation. Since
$R_{[\mu\nu]}$ depends only on the curl of $\mathcal A^\sigma$, this part
does not enter the torsional flow~\eqref{eq:torsional-flow}, consistently
with the vanishing static flip contribution in Ref.~\cite{r1}. The flip
representation~\eqref{eq:integral-hac-flip} is therefore understood as
an effective tetradic response to the prescribed mixed-chirality source,
rather than as a complete reconstruction of the independent axial
torsion field.

For a thin layer uniform parallel to the interface, we write
\begin{equation}\label{eq:flip-layer-source}
iS(z)=s_0\,\delta(z),
\end{equation}
where the real constant $s_0$ is the signed source strength
integrated across the layer. Integrating the static Yukawa kernel over the directions
parallel to the interface gives
\begin{equation}\label{eq:flip-layer-kernel}
g_{M_T}(z-z')
=\int_{\mathbb R^2}
G_{M_T}\!\left(\sqrt{|\boldsymbol{\rho}|^2+(z-z')^2}\right)
\,d^2\boldsymbol{\rho}
=\frac{e^{-M_T|z-z'|}}{2M_T}.
\end{equation}
Here $\boldsymbol{\rho}$ is the separation parallel to the plane.
The inner convolution is then
\begin{equation}\label{eq:flip-layer-response}
U(z)=\int_{\mathbb R}g_{M_T}(z-z')\,iS(z')\,dz'
=\frac{s_0}{2M_T}e^{-M_T|z|}.
\end{equation}
Its magnitude is largest at the interface and decays on both
sides over the screening length $M_T^{-1}$.

For the flip component $h_{[xy]}^{(\mathrm{flip})}$ of
Eq.~\eqref{eq:integral-hac-flip}, the factor $\epsilon_{byx}$
selects $b=z$. With $\epsilon_{xyz}=+1$, the prefactor
$-\epsilon_{zyx}/3$ is $+1/3$. We write the real profile
$q_{[xy]}^{(\mathrm{flip})}$. Integrating along a
straight path normal to the interface, with reference value zero at
$z=0$, gives
\begin{equation}\label{eq:flip-layer-tetrad}
q_{[xy]}^{(\mathrm{flip})}(z)
=\frac13\int_0^z U(u)\,du
=\frac{s_0}{6M_T^2}\operatorname{sgn}(z)
 \bigl(1-e^{-M_T|z|}\bigr).
\end{equation}
The reconstructed component is continuous at the interface.
Its derivative decays on the Yukawa scale, and the component
itself approaches a constant on each side. The difference is
\begin{equation}\label{eq:flip-layer-offset}
q_{[xy]}^{(\mathrm{flip})}(+\infty)
-q_{[xy]}^{(\mathrm{flip})}(-\infty)
=\frac{s_0}{3M_T^2}.
\end{equation}
This difference is the total rotation angle of the spatial triad
about the normal to the layer, accumulated across the interface.
For the planar source, $U$ depends only on $z$, so the component
$q_{[xy]}^{(\mathrm{flip})}$ is independent of the path, whereas the
components $q_{[yz]}^{(\mathrm{flip})}$ and $q_{[zx]}^{(\mathrm{flip})}$
are not. A M\"obius-like half twist would correspond to a rotation by
$\pi$, beyond the linear order used here.
If the source oscillates slowly on the screening scale,
$\omega\ll M_T$, these expressions hold at each instant with
$s_0\to s_0(t)$, and the difference oscillates with the source.
At fixed $M_T$, a change $\delta s_0$ therefore changes the
difference by $\delta s_0/(3M_T^2)$. Reversing the source
reverses the whole profile. The planar calculation gives the
response to a specified interface source. Determining that
source from fermion propagation in the toroidal field, and
testing the proposed spinorial holonomy, require a separate
transport calculation.

\paragraph{From helicity conversion to the interface response.}
The common ingredient of the two channels is the spin--torsion
interaction~\eqref{eq:Lint-A}: the potential that determines the
tetrad response of Section~\ref{sec:torus} is the same that enters
the spinor Hamiltonian~\eqref{eq:fourlevel}. The axial field
can change the helicity composition of a fermion without a first-order
deformation of the metric; for a Majorana neutrino, opposite helicities
correspond to different weak-interaction channels.
The fermion mass couples the chiral amplitudes. In a uniform
temporal background, $A^0$ changes their relative weights in
the positive-energy eigenstate, for which $S=0$. In a weak
purely transverse background, helicity conversion can generate
the oscillating source in Eq.~\eqref{eq:transverse-flip-source}.
At fixed background, fluctuations in the relative populations
of the two initial helicities can modulate this source.

For a prescribed layer, static or oscillating slowly on the
scale $M_T^{-1}$, Eq.~\eqref{eq:flip-layer-offset} gives the
difference in $q_{[xy]}^{(\mathrm{flip})}$ across the interface.
By Eq.~\eqref{eq:flip-drives-A0}, a source at a toroidal interface
requires an oscillating chiral imbalance. The remaining question is
whether propagation in the spatially varying Yukawa-screened axial
field $\mathcal A^\sigma$ produces such an oscillation, with a
nonzero integrated strength $s_0(t)$ at the interface between the
two domains.

\begin{table}[htbp]
\centering\small
\begin{tabular}{@{}p{0.35\linewidth}p{0.27\linewidth}p{0.29\linewidth}@{}}
\toprule
Configuration & Mixed source $S$ & Flip response \\
\midrule
Positive-energy eigenstate in a uniform temporal field
& $S=0$, including at equal chiral weights
& The sourced contribution vanishes \\

Initially definite helicity in a weak transverse field
& Oscillatory, with sign set by the initial helicity
& The source enters the Green convolution \\

Prescribed planar layer, static or with $\omega\ll M_T$
& $iS(z)=s_0\delta(z)$
& Asymptotic difference $s_0/(3M_T^2)$ \\
\bottomrule
\end{tabular}
\caption{Mixed-chirality source and antisymmetric tetrad response
in the configurations considered in Section~\ref{sec:fluct}.}
\label{tab:flip-response}
\end{table}
\section{Conclusions}

Starting from the Dirac Lagrangian in a first-order formulation,
we expressed the transverse part of the axial torsion and
contorsion through the antisymmetric tetrad perturbation, so that the torsional part of the
modified Ricci flow, Eq.~\eqref{eq:torsional-flow-h}, is written in
terms of it. Its fixed points require the curl of the screened axial
potential to vanish, Eq.~\eqref{eq:fixed-point}. The longitudinal
part, sourced by the mixed bilinear, is independent of the tetrad and
does not enter the flow. Within the
Ricci-flow framework of Refs.~\cite{r1,r2} and the integrated
formulation of Ref.~\cite{r3}, we then developed three Green-function
representations for the antisymmetric tetrad sector. The inner
convolution gives the response to the spinorial source, and the
outer integration reconstructs the tetrad contribution. The chiral
current difference sources the first two representations, while
the mixed bilinear $S$ enters the flip integral
\eqref{eq:integral-hac-flip}.

For a stationary source, the two current integrals contain the same
vector potential $A^k$, Eq.~\eqref{eq:vector-potential}. In physical
time, Eq.~\eqref{eq:triad-rotation} gives a rotation of the spatial
triad at a constant rate set by $A^k$. The toroidal configuration of Section~\ref{sec:torus} is the
stationary configuration of the Ricci flow obtained in
Ref.~\cite{r1}, modelled by a finite poloidal torus. Its magnetic-type dipole vanishes
and its toroidal moment is nonzero. In the unscreened limit,
Eq.~\eqref{eq:Phi} gives the circulation in terms of the signed
linking number, the current strength, and the torus radii.
Equation~\eqref{eq:Phi-thin} relates this circulation to the
toroidal moment in the thin-torus limit. With Yukawa screening,
the curl can extend outside the tube, and contours with the
same linking number can have different circulations. In the
unscreened limit the exterior of the tube satisfies the fixed-point
condition, so that the stationary configuration carries a nonzero
vector potential where the curvature vanishes, as anticipated in
Ref.~\cite{r1}.

In the spinor calculation, the fermion mass $m$ couples the two
chiral components. A uniform temporal axial field changes their
relative weights at fixed helicity. The mixed source $S$ vanishes
for the positive-energy eigenstate considered here, including
at equal chiral weights. A weak, uniform, purely
transverse axial field can instead induce helicity oscillations,
as in Eq.~\eqref{eq:flipfreq}, with an oscillating mixed source
given by Eq.~\eqref{eq:transverse-flip-source}. By
Eq.~\eqref{eq:flip-drives-A0}, the flip potential drives the temporal
axial potential. A static toroidal domain therefore carries no flip
source, whereas a configuration that is stationary in the sense of
the flow, with an oscillating chiral imbalance, can.

The flip integral defines a scalar potential, which twists the
spatial triad along one direction. To evaluate it explicitly, we
considered a prescribed mixed source concentrated on a thin planar
layer, static or oscillating slowly on the scale $M_T^{-1}$. The inner Yukawa convolution gives a response localized
near the interface on the scale $M_T^{-1}$, where $M_T$ is the
screening mass of the torsional sector. The outer integration
along the normal gives a continuous profile of
$q_{[xy]}^{(\mathrm{flip})}=i\,h_{[xy]}^{(\mathrm{flip})}$. Its two asymptotic values differ by
$s_0/(3M_T^2)$, where $s_0$ is the
signed source strength integrated across the layer
[Eq.~\eqref{eq:flip-layer-offset}]. This difference is the rotation
angle of the triad about the normal across the interface. At fixed $M_T$, fluctuations
of $s_0$ therefore give proportional fluctuations of this
difference.

The resulting chain is
$J_A^\mu\to\mathcal A^\sigma\to$ spinor dynamics $\to S\to U\to
h_{[\mu\nu]}^{(\mathrm{flip})}$. The axial potential does not produce
the flip by itself: it provides the transverse field in which the
helicity can oscillate, and by Eq.~\eqref{eq:flip-drives-A0} a
source arises only when the chiral imbalance oscillates.
The planar calculation gives the tetrad response to an assigned
interface source. To determine the source at a toroidal interface,
one must solve the propagation of the spinor in the spatially
varying axial field. The proposed M\"obius-like structure can then
be tested by computing the holonomy of the internal spinorial
frame along a closed contour.


\begin{thebibliography}{99}\setlength{\itemsep}{1pt}\setlength{\parsep}{0pt}
\bibitem{r1} E.~Varani, Geometric evolution of torsional curvature from Majorana neutrinos: a Ricci
flow approach, \emph{Transnational J.\ Math.\ Anal.} \textbf{13}(1) (2025) 53--104.
\bibitem{r2} E.~Varani, Topologically stabilized torsion in weak-field gravity: a Ricci-flow
framework, \emph{Adv.\ Theor.\ Comput.\ Phys.} \textbf{9}(3) (2026) 1--9; arXiv:2511.05635.
\bibitem{r3} E.~Varani, Torsional effects in the coupling between gravity and spinors: Yukawa
gravity, arXiv:2505.17158v2 (2025).
\bibitem{varani2025} E.~Varani, Torsion evidence and effects in spinor framework, in: F.~Bulnes (ed.), \emph{Notes of Seminar of Advanced Functional Analysis and Integral Transforms: Acta Matematica}, Cambridge Scholars Publishing, Newcastle upon Tyne, 2025, pp.~1--22.
\bibitem{bulnes2013} F.~Bulnes, Penrose Transform on Induced $D_{G/H}$-Modules and Their Moduli Stacks in the Field Theory, \emph{Advances in Pure Mathematics} \textbf{3}(2) (2013) 246--253. DOI: \href{https://doi.org/10.4236/apm.2013.32035}{10.4236/apm.2013.32035}.
\bibitem{bulnesgoborov2014} F.~Bulnes and R.~Goborov, Integral Geometry and Complex Space-Time Cohomology in Field Theory, \emph{Pure and Applied Mathematics Journal} \textbf{3}(6-2) (2014) 30--37. DOI: \href{https://doi.org/10.11648/j.pamj.s.2014030602.16}{10.11648/j.pamj.s.2014030602.16}.
\bibitem{bulnes2026} F.~Bulnes, Correspondence between hypercohomology groups to solutions as H-states with Hodge decomposition of a connection, \emph{Int.\ J.\ Adv.\ Appl.\ Math.\ Mech.} \textbf{14}(1) (2026) 64--71.
\bibitem{bulnes2022} F.~Bulnes, \emph{Deep Study of the Universe through Torsion}, Cambridge Scholars Publishing, Newcastle upon Tyne, 2022.
\bibitem{hamilton} R.~S.~Hamilton, Three-manifolds with positive Ricci curvature,
\emph{J.\ Differential Geom.} \textbf{17} (1982) 255--306.
\bibitem{ost2014} Yu.~N.~Obukhov, A.~J.~Silenko and O.~V.~Teryaev, Spin-torsion coupling and gravitational moments of Dirac fermions: theory and experimental bounds, \emph{Phys.\ Rev.\ D} \textbf{90} (2014) 124068.
\bibitem{hehl} F.~W.~Hehl, P.~von der Heyde, G.~D.~Kerlick and J.~M.~Nester, General relativity with
spin and torsion, \emph{Rev.\ Mod.\ Phys.} \textbf{48} (1976) 393--416.
\bibitem{shapiro} I.~L.~Shapiro, Physical aspects of the space-time torsion, \emph{Phys.\ Rep.}
\textbf{357} (2002) 113--213.
\bibitem{hop2013} F.~W.~Hehl, Yu.~N.~Obukhov and D.~Puetzfeld, On Poincar\'e gauge theory of gravity, its equations of motion, and Gravity Probe B, \emph{Phys.\ Lett.\ A} \textbf{377} (2013) 1775--1781.
\bibitem{lt} J.~Lense and H.~Thirring, \"Uber den Einflu\ss{} der Eigenrotation der Zentralk\"orper auf die Bewegung der Planeten und Monde nach der Einsteinschen Gravitationstheorie, \emph{Phys.\ Z.} \textbf{19} (1918) 156--163.
\bibitem{dubovik} V.~M.~Dubovik and V.~V.~Tugushev, Toroid moments in electrodynamics and solid-state
physics, \emph{Phys.\ Rep.} \textbf{187} (1990) 145--202.
\bibitem{ab} Y.~Aharonov and D.~Bohm, Significance of electromagnetic potentials in the quantum
theory, \emph{Phys.\ Rev.} \textbf{115} (1959) 485--491.
\bibitem{ck1997} D.~Colladay and V.~A.~Kosteleck\'y, CPT violation and the standard model, \emph{Phys.\ Rev.\ D} \textbf{55} (1997) 6760--6774.
\bibitem{cils1999} S.~Capozziello, G.~Iovane, G.~Lambiase and C.~Stornaiolo, Fermion helicity flip induced by torsion field, \emph{Europhys.\ Lett.} \textbf{46} (1999) 710--715.
\bibitem{dsg1981} V.~De Sabbata and M.~Gasperini, Neutrino oscillations in the presence of torsion, \emph{Nuovo Cimento A} \textbf{65} (1981) 479--500.
\end{thebibliography}
\end{document}